\documentclass{cernrep}
\usepackage[colorlinks=true, linkcolor=black, citecolor=black, filecolor=black, urlcolor=blue]{hyperref}
\usepackage{fancyhdr}
\fancyhfoffset{4 mm}
\fancypagestyle{ARTTITLE}{%
\fancyhf{} % clear all header and footer fields
\lhead{\footnotesize{Proceedings of the 2019 CERN--Accelerator--School course on
\it{ High Gradient Wakefield Accelerators}, Sesimbra, (Portugal)}}
\lfoot{Available online at \url{https://cas.web.cern.ch/previous-schools}}
\rfoot{\thepage\hspace*{3mm}}

   }
\begin{document}

\title{Staging of High-Gradient Wakefield Accelerators}
\author{C. A. Lindstr{\o}m}
\institute{Deutsches Elektronen-Synchrotron DESY, Hamburg, Germany}

\begin{abstract}
    Accelerating particles to high energies with a high-gradient wakefield accelerator may require use of multiple stages. Coupling beams from one stage to another can be difficult due to high divergence and non-negligible energy spreads. We review the challenges, technical requirements and currently proposed methods for solving the staging problem.
\end{abstract}

\keywords{Staging; wakefield accelerators; plasma wakefields; chromaticity.}
\maketitle
\thispagestyle{ARTTITLE}
\section{Introduction}
Accelerating particles to high energy in a compact space is the main motivation for high-gradient wakefield accelerator research. What counts as \textit{high energy}, however, depends on the context---free-electron lasers \cite{MadeyFELJAP1971} will require energies around 1--10~GeV, whereas linear colliders \cite{ILCTDR2013,CLICCDR2013} will require 100~GeV or more. The energy gain in a single wakefield accelerator stage \cite{TajimaPRL1979,ChenPRL1985,RuthPA1985,GaiPRL1988} is ultimately limited by the energy stored in its driver, which is typically around 1--100~Joules. While a single stage may be enough to drive a free-electron laser, it will not be sufficient to drive a linear collider. In this case we need to combine the~energy of several individual drivers, by distributing these drivers across many separately driven stages chained together---also known as \textit{staging}.

Staging is still largely an unsolved problem with no universal solution, although good progress has been made in recent years. With only one dedicated experimental result to date \cite{SteinkeNature2016}, there are still many problems to be solved on a conceptual level. In this Chapter, we will explore why staging is so challenging (Section~\ref{sec:Challenges}), the detailed technical requirements (Section~\ref{sec:Requirements}), and finally some currently proposed methods to get there (Section~\ref{sec:ProposedMethods}).

\section{The Staging Problem}
\label{sec:Challenges}
On the face of it, staging sounds like a rather simple way to reach higher energies. After all, we have been doing it for a century with conventional accelerating cavities: just put one after the other! However, in high-gradient wakefield accelerators this turns out to be anything but simple---subtle but fundamental effects complicate matters immensely. Before delving into the details of how to do staging, we need to gain an understanding of why it is so difficult in the first place.

Here is the problem: to accelerate particles with high gradients, strong focusing is required to maintain stable acceleration. This strong focusing results in high-divergence beams in the space between the stages---making it difficult to preserve the beam quality of finite-energy-spread bunches. This section will attempt to answer three questions: Why do the stages need to be separated in space? Why is strong focusing necessary within stages? And why are high-divergence beams so difficult to handle?

\subsection{In- and out-coupling of the driver}
\label{sec:InOutCoupling}
A defining feature of wakefield accelerators is the co-propagating driver---the source of energy for the~accelerating particles. When the driver has been depleted of its energy, it must be swapped out for a fresh one. This extraction and re-injection process will inevitably disrupt the wakefield for some time before the next wakefield can be set up. In practice, this corresponds to a gap between the stages.

Methods for swapping out the driver will depend on the type of driver being used---lasers or charged particle beams. Two devices are available for making the driver collinear with the accelerating particle beam: mirrors and magnetic dipoles. Clearly, mirrors can only be used for laser-drivers, whereas dipoles can be used in both laser- and beam-driven accelerators.

\begin{figure}[ht]
	%\centering\includegraphics[width=0.9\linewidth]{BELLAstaging.pdf}
	\centering
	\href{https://www.nature.com/articles/nature16525/figures/1}{\includegraphics[width=0.93\linewidth]{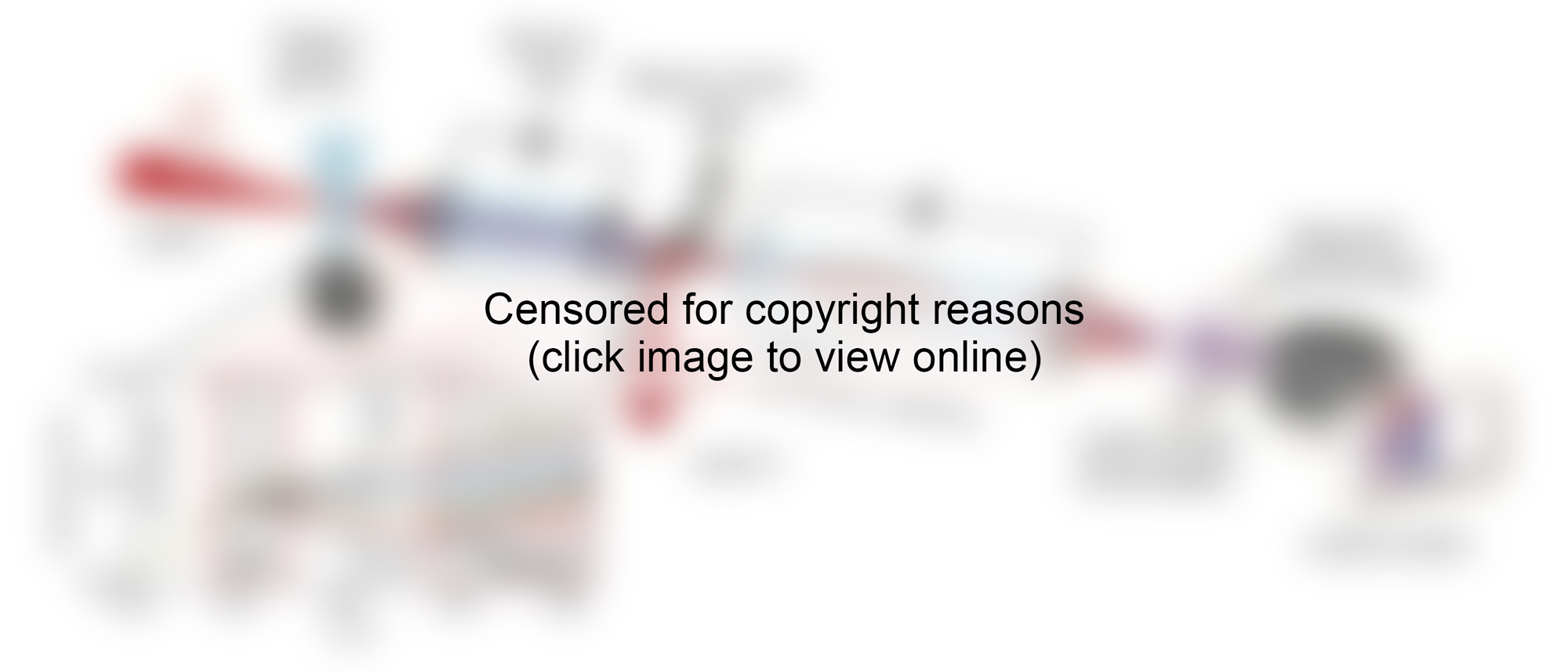}}
	\caption{The BELLA staging experiment performed at LBNL. Electrons from a laser wakefield accelerator stage were focused into a second stage using a plasma lens. Simultaneously, a second laser driver was coupled in using a plasma mirror. Electrons with a central energy of 120~MeV were boosted in the second stage by up to 100~MeV. A large energy spread (60\% full width at half-maximum) resulted in a large chromaticity and consequently a~low charge-coupling efficiency between the two stages (about 3.5\%). Source: S.~Steinke \textit{et al.}, Nature \textbf{530}, 190 (2016) \cite{SteinkeNature2016}.}
    \label{fig:BELLAstaging}
\end{figure}

Mirrors are ideal in that they do not require much longitudinal space along the beam axis---the~laser pulse is coupled in and out transversely. A holed mirror can be used to let the beam through undisturbed. However, if placed very close to the stage, the laser intensity is typically too high: a normal optical mirror would burn immediately. One solution is to place the mirror further away from the stage, resulting in no burning but taking up more space. Alternatively, one can use a so-called \textit{plasma mirror} \cite{ThauryNatPhys2007}---a~thin foil which vaporizes on contact, turning into a solid-density plasma that reflects the laser. Such plasma mirrors were successfully used in the first and currently only staging experiment, performed at the Lawrence Berkeley National Lab \cite{SteinkeNature2016} (see Fig.~\ref{fig:BELLAstaging}). However, preserving the emittance of a beam passing through a plasma mirror can be difficult, due to effects such as beam filamentation \cite{AllenPRL2012,RajArxiv2019} (for small beams) and Coulomb scattering \cite{WiedemannSpringer2007,KirbyPAC2007} (for large beams).

%--- Figure 1: BELLA Experiment ---%

Magnetic dipoles bend the path of charged particles, and can therefore be used to merge/separate a particle beam and a laser beam, or merge/separate two different particle beams. In the latter case, we would ideally use a fast kicker magnet \cite{BarnesCAS2011}---injecting and extracting the driver bunch while leaving the accelerated bunch undisturbed. However, no kicker exists that can separate bunches at the necessary femtosecond-to-picosecond timescales (the fastest kickers have a nanosecond rise time \cite{ArntzIEEE2007}). Therefore, currently the only way to separate two beams is by energy---i.e.,~by using beam drivers with lower or higher energy than the accelerated bunch. Effectively, this also applies to laser drivers, which have no charge and therefore behave as if they have a higher (infinite) energy. Unfortunately, the~merging/separation process introduces large quality-degrading dispersion to both accelerating beams and beam drivers, which must be carefully cancelled---taking up valuable space along the beamline.

Whichever method is chosen for in- and out-coupling of the driver, it will introduce a separation between the stages---likely between 0.1~and~10~m per stage. This will not only reduce the effective accelerating gradient of a multi-stage accelerator, but also leads to trouble when combined with the~strong focusing within each stage.

\subsection{The need for strong focusing}
Increasing the accelerating gradient generally requires decreasing the transverse dimensions of the accelerator cavity. This is particularly true for plasma wakefield accelerators, where the characteristic accelerating field $E_z$ is directly linked to the inverse of the characteristic length scale of the plasma wake---the plasma skin-depth $1/k_p$:
\begin{equation}
    \frac{E_z}{m_e c^2} \simeq k_p,
\end{equation}
where $k_p = \sqrt{n e^2/ m_e \epsilon_0 c^2}$ is the plasma wavenumber, $n$ is the plasma density, $\epsilon_0$ and $c$ are the vacuum permittivity and light speed, and $m_e$ and $e$ are the electron mass and charge, respectively.

\begin{figure}[hb]
	\centering\includegraphics[width=0.9\linewidth]{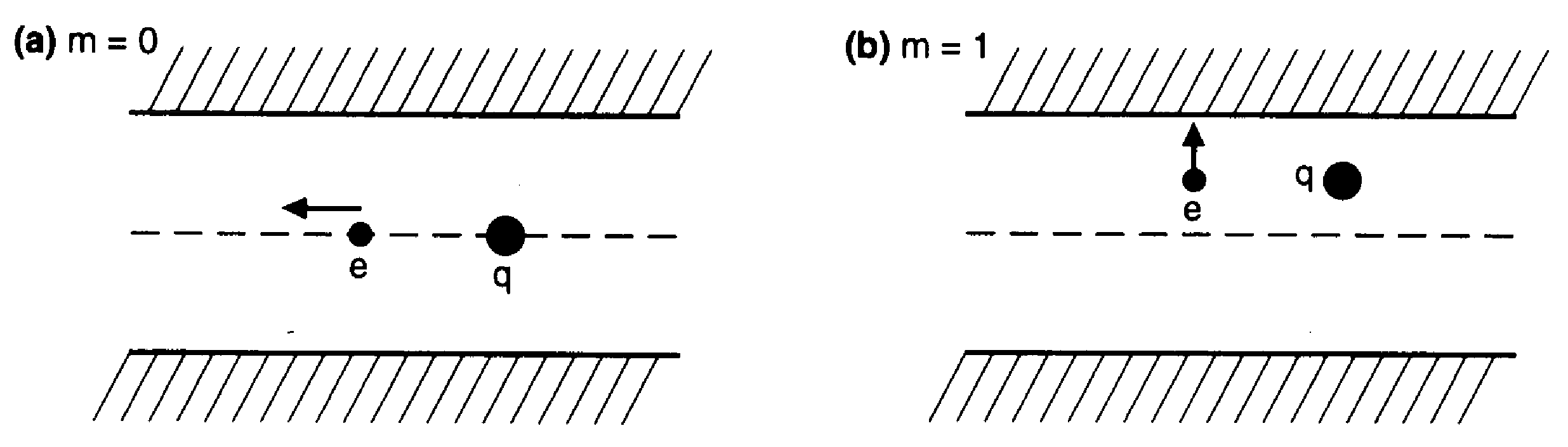}
	\caption{Two particle bunches in a generic accelerator structure, exhibiting both (a) a longitudinal wakefield of azimuthal mode $m=0$ and (b) a transverse wakefield of azimuthal mode $m=1$. Arrows indicate the direction of the force. Source: A.~W.~Chao,
    \textit{Physics of Collective Beam Instabilities in High-Energy Accelerators} (Wiley, New York, NY, 1993) \cite{ChaoBook1993}.}
    \label{fig:WakefieldModes}
\end{figure}

In general, while the longitudinal accelerating field makes up the fundamental mode of this wakefield, other modes will also be present \cite{ChaoBook1993}---in particular the transverse wakefield (see Fig.~\ref{fig:WakefieldModes}). This deflecting force scales with the transverse offset, so theoretically if the beam was perfectly aligned on axis, it would be unaffected. However, since the transverse wakefield always pulls the beam away from the axis, a positive feedback loop of larger offsets and larger wakefields is induced for even an infinitesimal initial offset. This instability was discovered in the SLAC linac back in 1960 \cite{KelliherNature1960}, where transverse wakefields from the beam pipe \cite{BaneAIP1996} caused parts of the beam to be lost. An asymmetry in a high-gradient wakefield accelerator will have the same effect.

%--- Figure: Wakefield modes ---%

The fundamental challenge is that when the characteristic cavity dimension ($a$) decreases, the~longitudinal wakefield increases (typically between $1/a$ and $1/a^2$), but the transverse wakefield grows even faster (typically between $1/a^3$ and $1/a^4$). This is dictated by the Panofsky-Wenzel theorem \cite{PanofskyWenzelRSI1956}, and is always true in the short range (i.e., directly behind the driver), where it is known as the \textit{short-range wake theorem} \cite{LebedevPRAB2017}
\begin{equation}
    \label{eq:ShortRangeWakeTheorem}
    \frac{W_x(z)}{\Delta x} = -\frac{2}{a^2} \int_0^z W_z(z') dz',
\end{equation}
where $W_x$ and $W_z$ are the transverse and longitudinal wakefields (force per charge), respectively, $\Delta x$ is the beam offset, and $z$ is the co-moving longitudinal coordinate.

The good news is that the transverse wakefield instability can be suppressed by introducing a~comparatively strong focusing force. If at a certain transverse offset, a particle observes a stronger focusing force toward the axis than a transverse wakefield toward the wall, the beam will remain stable. In conventional accelerators like the SLAC linac and CLIC, this focusing is done with quadrupoles \textit{interleaved} between the accelerator cavities. However, this is not sufficient for a high-gradient wakefield accelerator since the transverse wakefield is too strong---focusing must occur \textit{inside} the accelerating cavity. Fortunately, many wakefield accelerators provide intrinsically strong internal focusing. This is one of the main advantages of plasma wakefield accelerators, where an exposed ion column (in the case of a nonlinear wake) focuses the electron beam with a focusing strength (force per offset) of
\begin{equation}
    \label{eq:PlasmaFocusing}
    K = \frac{k_p^2}{2\gamma},
\end{equation}
where $\gamma$ is the relativistic Lorentz factor. Other wakefield accelerators, such as dielectric wakefield accelerators, plan to use strong external focusing (quadrupoles) \cite{LiPRSTAB2014}. Regardless of the focusing mechanism, in-situ focusing is generally required for stable operation of high-gradient wakefield accelerators.

Finally, while strong focusing is required, it can still lead to a so-called \textit{beam breakup instability} \cite{PanofskyRSI1968}---or similarly a \textit{hose instability} in a plasma accelerator \cite{WhittumPRL1991,HuangPRL2007}. This is caused by a resonance between different longitudinal slices of the bunch when oscillating in the focusing field. To avoid this effect the slices must be decohered to oscillate at different frequencies. This is normally done by giving the bunch a head-to-tail energy chirp (a longitudinally correlated energy spread)---a method known as \textit{BNS damping} after Balakin, Novokhatsky and Smirnov \cite{BalakinVLEPP1983}. The implication is that stable acceleration benefits from a non-zero energy spread \cite{MehrlingPRL2017}.

\subsection{Chromaticity}
Strong focusing results in small beam sizes, but more importantly, highly diverging beams. In terms of Courant-Snyder or Twiss parameters \cite{CourantSnyderAoP1957}, this means small beta functions. To avoid emittance growth in a stage, the beta function must be \textit{matched} to
\begin{equation}
    \beta_m = \frac{1}{\sqrt{K}},
\end{equation}
where $K$ is the focusing strength---in this case, the natural divergence of the beam is exactly countered by the focusing field such that the beta function (and hence the beam size) stays constant. For a plasma accelerator (Eq.~(\ref{eq:PlasmaFocusing}))
\begin{equation}
    \label{eq:MatchedBeta}
    \beta_m = \frac{\sqrt{2\gamma}}{k_p} = \sqrt{\frac{2 \epsilon_0 E}{n e^2}},
\end{equation}
which is typically on the mm-to-cm scale for beam energies $E$ on the GeV-level. Capturing and refocusing these rapidly diverging beams can be challenging \cite{AnticiJAP2012,MiglioratiPRSTAB2013}. High-gradient wakefield acceleration often results in non-negligible energy spread because of the rapidly changing (high-frequency) accelerating field structure---in plasmas the energy spread is often 1\% or more. While this may be good for BNS damping, these beams are difficult to capture and refocus without degrading the beam quality because the different energy slices cannot all be focused in the same way---an effect known as \textit{chromaticity} . 

The chromaticity of the beam focusing is typically defined in terms of the \textit{chromatic amplitude} \cite{ZyngierLAL1977, MontagueLEP1979}
\begin{equation}
    W = \sqrt{\left(\frac{\partial\alpha}{\partial\delta}-\frac{\alpha}{\beta}\frac{\partial\beta}{\partial\delta}\right)^2 +  \left(\frac{1}{\beta} \frac{\partial\beta}{\partial\delta}\right)^2},
\end{equation}
which measures (to first order) the combined mismatch of the Twiss parameters $\alpha$ and $\beta$, for a relative energy offset $\delta = \Delta E/E$. This chromatic amplitude can be related to the projected (energy-averaged) emittance growth via \cite{LindstromPRAB2016}
\begin{equation}
    \frac{\Delta \epsilon^2}{\epsilon_0^2} = W^2 \sigma_{\delta}^2 + \mathcal{O}(\sigma_{\delta}^4),
\end{equation}
expressed to lowest order in $\sigma_{\delta}$ (the relative rms energy spread). It is important to note that this projected emittance growth is not a ``true'' emittance growth, since the emittance of each energy slice is conserved---it can therefore in principle be reversed. However, when entering the next stage and observing strong focusing, such a phase space reversal is practically impossible.

%--- Figure: Chromaticity ---%
\begin{figure}[ht]
	\centering\includegraphics[width=0.95\linewidth]{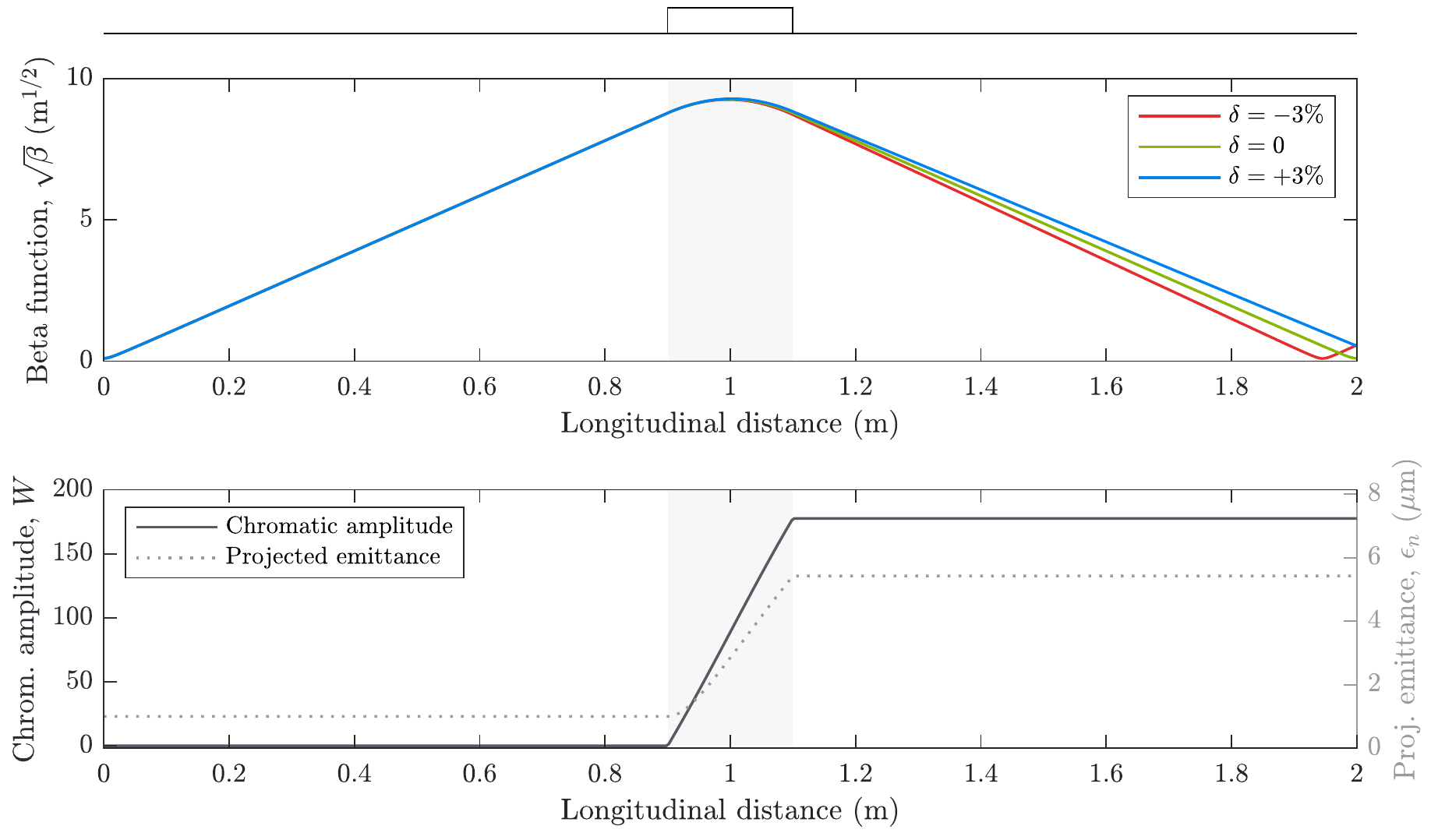}
	\caption{Example of emittance growth due to chromaticity. A 10~GeV beam with 3\%~rms energy spread diverges from a plasma accelerator of density $10^{16}$~cm$^{-3}$ ($\beta_m \approx 10$~mm). A simple beam optics lens captures and refocuses the beam into the next stage, which introduces significant chromaticity. As a result, the projected (energy-averaged) emittance increases by more than a factor 5.}
    \label{fig:Chromaticity}
\end{figure}

So how large do we expect the emittance growth to be? Consider a simple example of staging: a~beam of energy spread $\sigma_{\delta}$ exits a stage with a beta function $\beta_m$, then gets refocused by a (thin) magnetic optics lens after a drift of length $L$ (see Fig.~\ref{fig:Chromaticity} for illustration). In simple cases like this, the chromaticity added in the~lens is approximately $\Delta W = \beta Kl$, where $Kl$ is the integrated focusing strength and $\beta$ is the~beta function in the lens. To capture and refocus the beam each in a distance $L$, the focal length of the~lens must be $L/2$---resulting in an integrated focusing strength of $Kl = 2/L$. Outside the~stage, the~beam will diverge to a beta function $\beta \approx L^2/\beta_m$, assuming a small matched beta function (i.e., $\beta_m \ll L$). Putting it all together, we find that the chromaticity is $W \approx 2L/\beta_m$ and therefore the~projected emittance growth will be approximately
\begin{equation}
    \frac{\Delta \epsilon^2}{\epsilon_0^2} \approx \frac{4 L^2}{\beta_m^2} \sigma_{\delta}^2.
\end{equation}
This sets strict limits for the acceptable energy spread. Take for instance a plasma accelerator stage at energy $E=10$~GeV with plasma density $n=10^{16}$~cm$^{-3}$ (giving $\beta_m=10$~mm), using a capture length of $L=1$~m, and limited to an emittance growth of 1\%---the maximum energy spread is only 0.07\% rms!

In conclusion, chromaticity places severe constraints on the staging of high-gradient wakefield accelerators if left uncorrected.

\pagebreak

\section{Technical Requirements}
\label{sec:Requirements}
While chromaticity is perhaps the biggest challenge, there are also many other considerations to keep in mind when designing a coherent staging scheme. In this section, we will review some of the most important requirements in detail.

\subsection{Emittance preservation}
Delivering low-emittance beams is of prime importance for most high-energy accelerator applications. In a linear collider, the \textit{luminosity}---or event rate---is given by \cite{CLICCDR2013}
\begin{equation}
    \label{eq:Luminosity}
    \mathcal{L} = H_D \frac{N^2 f \gamma}{4 \pi \sqrt{\beta_x\epsilon_{nx}}\sqrt{\beta_y\epsilon_{ny}}},
\end{equation}
where $N$ is the number of bunch particles, $f$ is the collision frequency, $H_D$ is a numerical factor, $\beta_x$ and $\beta_y$ are the interaction point beta functions, and $\epsilon_{ny}$ and $\epsilon_{ny}$ are the normalized transverse emittances. Similarly, in a free-electron laser, the lasing power is determined by the \textit{6D brightness} \cite{LejeuneACPO1980}
\begin{equation}
    \label{eq:Brightness6D}
    B_{\textmd{6D}} = \frac{N}{\epsilon_{nx}\epsilon_{ny}\epsilon_{nz}},
\end{equation}
where $\epsilon_{nz}$ is the normalized emittance of the longitudinal phase space. Equations \ref{eq:Luminosity} and \ref{eq:Brightness6D} indicate that low emittance is crucial---both applications require normalized transverse emittances of 1~mm~mrad or lower. Technically, conservation of charge ($N$) is equally important, and in practice we will require close to 100\% charge-coupling efficiency between stages.

Producing these low-emittance bunches is routinely done using photocathodes \cite{PierceRSI1980} or even plasma injection techniques \cite{LeemansNatPhys2006,BuckPRL2013,DengNatPhys2019}. However, preserving the emittance through a large number of stages sets very stringent limits on the emittance growth per stage. It can be useful to imagine an \textit{emittance budget}, where each stage only gets to contribute a certain emittance growth to the final emittance. For wakefield accelerators with many stages, this will typically be on the 0.1~mm~mrad-level or less per stage.

\subsubsection{Matching}
\label{sec:Matching}

%--- Figure: Mismatching ---%
\begin{figure}[b]
	\centering\includegraphics[width=0.9\linewidth]{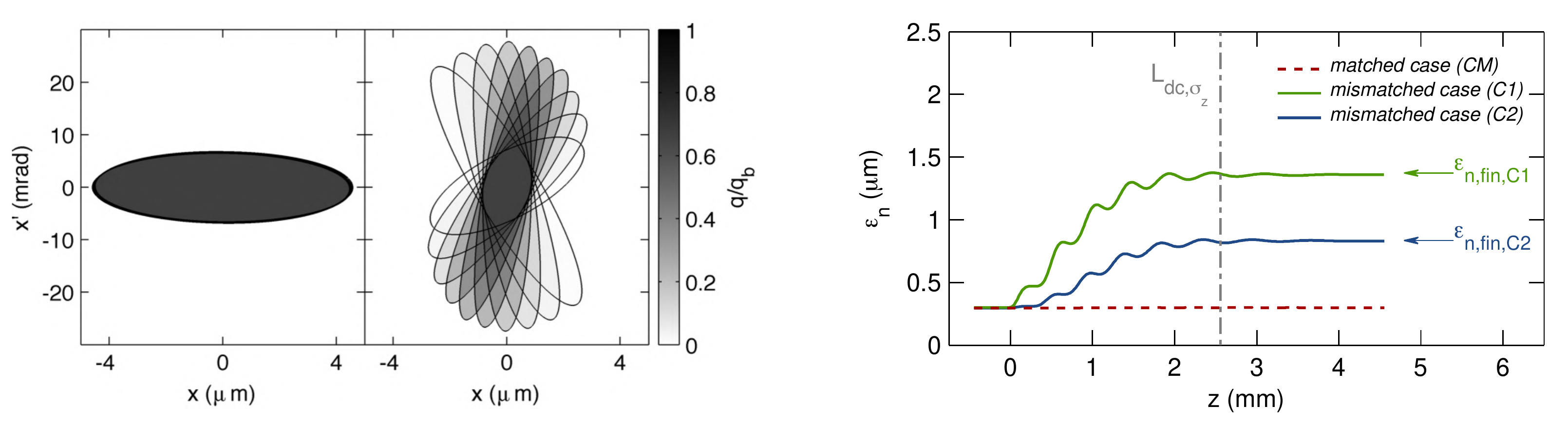}
	\caption{Mismatching of a beam with finite energy spread. The initial phase space (left panel) is mismatched to the~focusing channel, which leads to a smearing in phase space due to the different rates of rotation. This effect is seen to saturate at some point (right panel), when the beam has been fully smeared in phase space, leading to emittance growth. Source: T.~Mehrling \textit{et al.}, Phys.~Rev.~ST Accel.~Beams \textbf{15}, 111303 (2012) \cite{MehrlingPRAB2012} (CC BY 3.0).}
    \label{fig:Mismatching}
\end{figure}

Mismatching bunches with a finite energy spread leads to emittance growth because the phase space ellipses of different energy-slices rotate at different rates---eventually smearing out to a larger area. The~end result is a matched beam with a larger emittance (see Fig.~\ref{fig:Mismatching} for illustration). This process stops (or \textit{saturates}) after a distance $L_{\textmd{sat}} \approx \beta_m/\sigma_{\delta}$, after which the saturated emittance will be \cite{MehrlingPRAB2012}
\begin{equation}
    \label{eq:Mismatching}
    \frac{\epsilon_{\textmd{sat}}}{\epsilon} = \frac{1}{2}\left( (1+\alpha^2)\frac{\beta_m}{\beta} + \frac{\beta}{\beta_m} \right),
\end{equation}
where $\alpha$ and $\beta$ are the Twiss parameters at the plasma entrance (assuming a flat top density profile). As seen in Eq.~(\ref{eq:Mismatching}), the only way to avoid emittance growth from mismatching is to ensure that $\beta = \beta_m$ and $\alpha = 0$.

Technically, this is also why chromaticity is problematic---if the central energy slice is matched, but other energy slices are not, this will result in an emittance growth. Hence, to avoid mismatching, chromaticity between stages must be cancelled.

\subsubsection{Dispersion cancellation}
\label{sec:Dispersion}
When dipoles are used for in- and out-coupling of drivers, a correlation between energy and position---\textit{dispersion}---is intentionally introduced to separate beams of different energy. This also disperses the~accelerating beam if it has a nonzero energy spread, introducing a projected emittance growth. We can estimate this emittance growth in a stage from an uncorrected (first-order) dispersion to be
\begin{equation}
    \label{eq:Dispersion}
    \Delta\epsilon_D \approx \frac{1}{2}\left( \frac{D_x^2}{\beta_m} + \beta_m D_{x'}^2 \right) \sigma_{\delta}^2,
\end{equation}
where $D_x$ is the dispersion, $D_{x'}$ is the dispersion prime (i.e., energy--angle correlation), and $\beta_m$ is the~matched beta function in the stage. Using this relation, we can obtain approximate limits for (first-order) dispersion: $D_x \ll \sqrt{2 \epsilon \beta_m}/\sigma_{\delta}$ and $D_{x'} \ll \sqrt{2 \epsilon /\beta_m}/\sigma_{\delta}$, where $\epsilon$ is the geometric emittance of the beam. As an example, a 1~GeV beam with 1\% energy spread and 1~mm~mrad normalized emittance staged between plasma accelerators of density 10$^{16}$~cm$^{-3}$ will require dispersion and dispersion-prime cancellation to much better than 0.18~mm and 55~mrad, respectively---this can be quite challenging. We should also note that given the large dispersion often introduced in strong dipoles, we may also need to consider higher-order dispersion.

Moreover, dispersion can cause additional problems beyond just an increased projected emittance. If the longitudinal phase space of the bunch has a correlation---as it often does---a dispersion implies that the bunch has a tilt and/or a curvature. Such an asymmetry can seed a beam breakup \cite{PanofskyRSI1968} or hosing instability \cite{WhittumPRL1991,HuangPRL2007}, which can lead to more severe emittance growth.

\subsection{Isochronicity}
Inside a stage, one rarely has to worry about changes to the bunch length---effectively, it is conserved. However, outside of the stage there are multiple ways in which the bunch can be lengthened or compressed. This can be detrimental to beam loading and energy-spread conservation \cite{KatsouleasPRA1986,TzoufrasPRL2008}.

If dipoles are used to separate a driver and an accelerating bunch, perhaps in the form of a chicane, particles of different energy may travel different distances before arriving at the next stage---this~is the~idea behind a bunch compressor. In technical terms, we talk about the $R_{56}$ matrix element of the~accelerator lattice, also known as the \textit{longitudinal dispersion} ($R_{16}$ and $R_{36}$ are the horizontal and vertical dispersions). For the bunch length to be conserved during staging, we will require that $R_{56} = 0$, in which case the lattice is called \textit{isochronous}. We can place a limit on this condition for a bunch of a given energy spread $\sigma_{\delta}$ and bunch length $\sigma_z$:
\begin{equation}
    \label{eq:R56}
    \left| R_{56} \right| \ll \frac{\sigma_z}{\sigma_\delta}.
\end{equation}

\begin{figure}[t]
	\centering\includegraphics[width=0.65\linewidth]{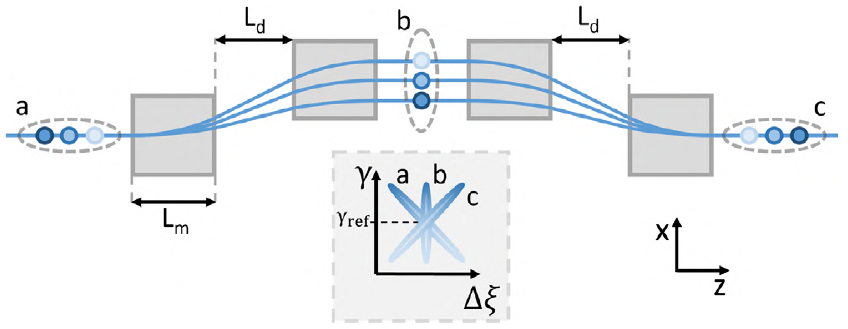}
	\caption{Longitudinal phase space reversal in a chicane with a fine-tuned $R_{56}$, which can be used to significantly reduce the~energy spread by compensating the chirp from one stage in the following stage. Source: A.~Ferran Pousa \textit{et al.}, Phys.~Rev.~Lett.~\textbf{123}, 054801 (2019) \cite{FerranPousaPRL2019} (reproduced with permission).}
    \label{fig:LPSreversal}
\end{figure}

In special cases, it can be beneficial to not cancel $R_{56}$ completely---especially if combined with chirped accelerator stages. If one stage produces a chirped bunch (e.g., the front particles have lower energy), a carefully tuned $R_{56}$ can overcompress the bunch in such a way that the bunch length is conserved, but the longitudinal phase space is flipped. In this case the next stage will exactly dechirp the~bunch to give it a significantly lower energy spread (see Fig.~\ref{fig:LPSreversal}) \cite{FerranPousaPRL2019}.

%--- Figure: Bunch chirp compensation ---%

A different, but related problem can occur even if the longitudinal dispersion is cancelled and the~energy spread is negligible. If the bunch is very short ($\mu$m-scale) and the divergence is large---as is often the case in laser plasma accelerators---particles with a large angle will travel further before being focused back to a small beam size (related to the $R_{25}$ and $R_{45}$ matrix elements). To avoid bunch lengthening, the distance to the first/last optic before and after a stage is therefore restricted by
\begin{equation}
    \label{eq:Synchronicity}
    \Delta L \approx \frac{1}{2}\sigma_{x'}^2 L \ll \sigma_z,
\end{equation}
where $\Delta L$ is the path length difference for particles at an angle of $\sigma_{x'}$ (i.e., the rms divergence), and $L$ is the distance to the optic. Normally, this is only a problem for lower energy bunches (sub-GeV), because the matched beta function is smaller ($\sim\sqrt{\gamma}$) and the geometric emittance is higher ($\sim1/\gamma$)---both leading to higher divergence. For example, a 200~MeV bunch of length 1~$\mu$m and normalized emittance 1~mm~mrad exiting a laser plasma accelerator at density 10$^{17}$~cm$^{-3}$ will have a divergence of 2.3~mrad and therefore must according to Eq.~(\ref{eq:Synchronicity}) be captured in much less than 37~cm.

\subsection{Tolerances}
All of the above considerations have assumed that the staged accelerator is perfectly stable. This is, of course, not the case in practice---everything has a certain level of random jitter. Two particularly important tolerances for jitter are those related to the synchronization and the transverse misalignment between the driver and the accelerating beam.

\subsubsection{Synchronization}
High-gradient wakefield accelerators have high-frequency electromagnetic fields. For stable acceleration, the driver and the accelerating beam must be synchronized to within a small fraction of the wakefield period. Consider an accelerator with a wakefield that changes from 0 to $E_z$ in a time $1/\omega$---an error in the relative arrival time $\Delta t$ will result in a relative error of the accelerating gradient of
\begin{equation}
    \label{eq:Synchronization}
    \frac{\Delta E_z}{E_z} \approx \omega \Delta t.
\end{equation}
Random timing jitter therefore results in a corresponding energy jitter---an effective multi-shot energy spread. As an example, to maintain a 1\% energy stability in a plasma accelerator stage operating at density 10$^{17}$~cm$^{-3}$ (characteristic timescale $1/\omega_p = 177$~fs), one would need to synchronize the driver and the accelerating beam to better than 2~fs. This is a very challenging goal---current state-of-the-art techniques can provide synchronization jitter down to about 10~fs rms \cite{SchulzNatComms2015,ShallooCLF2015}. In addition, long-term timing drifts need to be measured and corrected for with a feedback system operating at the same timescales.

\subsubsection{Transverse misalignments}
Misalignment tolerances will also prove particularly challenging for high-gradient wakefield accelerators. The accelerating beam, the driver, and the accelerating structure all need to be well aligned throughout the full length of the accelerator. Fortunately, this problem is partially mitigated in a plasma accelerator since the driver defines the location of the accelerating structure---only the driver--accelerating beam offset matters. Assuming instabilities such as hosing and beam breakup can be mitigated, what level of emittance growth do we expect?  

\begin{figure}[t]
	\centering\includegraphics[width=0.95\linewidth]{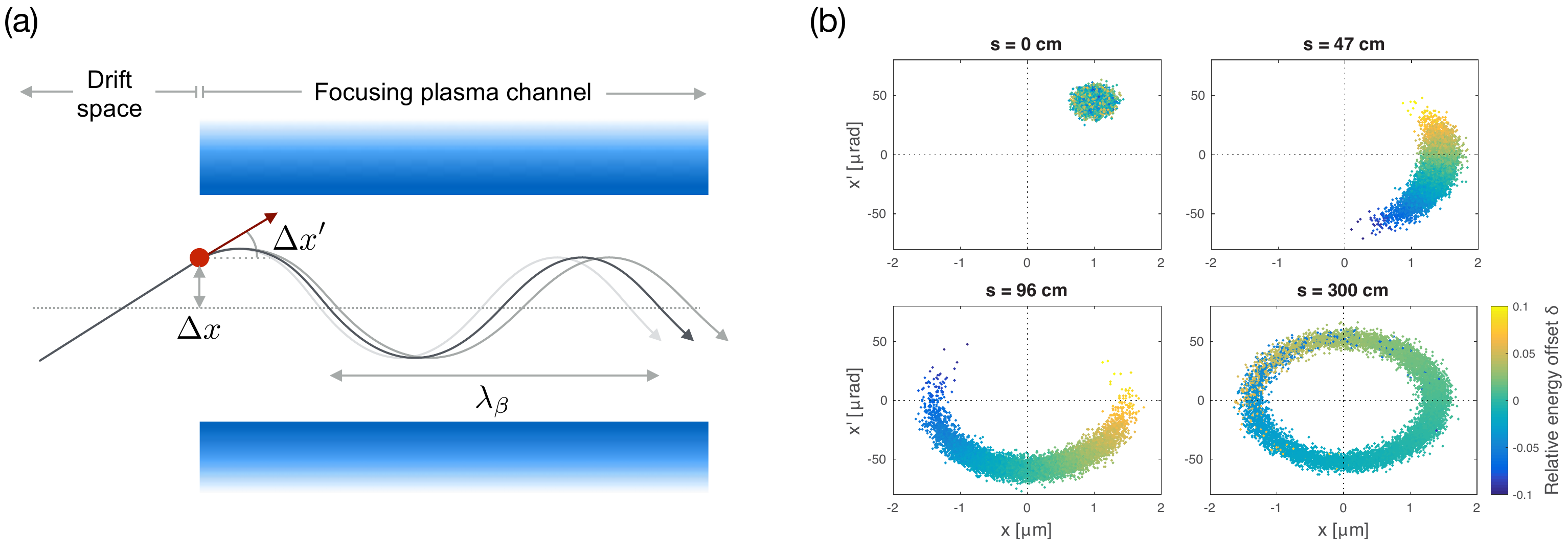}
	\caption{(a) Illustration of projected emittance growth due to misalignments in a strong-focusing channel. (b)~The~accelerating bunch starts with an offset and an angle, which due to a finite energy spread smears out to a larger-area ring in phase space. Source: C.~A.~Lindstr{\o}m \textit{et al.}, \textit{Proceedings of IPAC2016} (JACoW, Geneva, 2016), p.~2561 \cite{LindstromIPAC2016} (CC BY 3.0).}
    \label{fig:Misalignments}
\end{figure}

Similar to the cases of mismatching (see Section~\ref{sec:Matching}) and dispersion (see Section~\ref{sec:Dispersion}), beams with finite energy spread will see the centroids of different energy slices rotate in phase space at different rates---leading to a smearing in phase space. Consider a driver--accelerating beam pair with a relative position offset $\Delta x$ and an angle offset $\Delta x'$, propagating in a stage with a matched beta function $\beta_m$ (see Fig.~\ref{fig:Misalignments} for illustration). The projected emittance growth caused by such an offset will gradually increase along the accelerator, and then saturate at \cite{LindstromIPAC2016}
\begin{equation}
    \Delta\epsilon_j \approx \frac{1}{2}\left( \frac{\Delta x^2}{\beta_m} + \beta_m \Delta {x'}^2 \right).
\end{equation}
In simple terms, the driver and the accelerating beam must overlap well in phase space: $\Delta x \ll \sqrt{2 \beta_m\epsilon}$ and $\Delta x \ll \sqrt{2\epsilon/\beta_m}$, where $\epsilon$ is the geometric emittance of the accelerating beam. This is particularly challenging for small $\beta_m$ (i.e., high accelerating gradients) and lower emittances, as the beam is focused to the sub-$\mu$m-scale. If multiple stages are used, the emittance growth per stage is further constrained. For plasma accelerators with relevant gradients (GV/m-scale), the alignment tolerance is estimated to be around 10--50~nm \cite{AssmannNIMA1998, CheshkovPRAB2000,SchulteRoASaT2016}.

%--- Figure: Transverse misalignments ---%

\subsection{Synchrotron radiation}
As discussed in Section~\ref{sec:InOutCoupling}, one of the main methods to separate the driver and the accelerating beam is to disperse them using a magnetic dipole---possibly the only way for a beam-driven wakefield accelerator. When accelerating to higher energies---the objective of staging---we will naturally hit the same problem as we tried to avoid by using a linear accelerator (as opposed to a circular accelerator): \textit{synchrotron radiation}. For long bunches, particles emit \textit{incoherent} synchrotron radiaton (ISR), with an average power emitted per bunch \cite{LarmorPhilMag1897,Jackson1999}
\begin{equation}
    P_{\textmd{ISR}} = \frac{e^4}{6\pi \epsilon_0 m^2 c} N \gamma^2 B^2,
\end{equation}
where $m$ and $e$ are the particle mass and charge, $c$ and $\epsilon_0$ are the vacuum light speed and permittivity, $\gamma$ is the relativistic Lorentz factor, $B$ is the magnetic field, and $N$ is the number of particles.

However, high-gradient wakefield accelerators often operate with short bunches, in which case the~bunch will emit \textit{coherent} synchrotron radiation (CSR) \cite{SaldinNIMA1997,NovokhatskiICFA2012}. In the case of full coherence, the~electric fields of all particles add linearly such that the radiated power scales quadratically with particle number
\begin{equation}
    P_{\textmd{CSR}} = N P_{\textmd{ISR}} \sim N^2.
\end{equation}
The radiation is fully coherent only if the bunch length is less than $\sigma_{\textmd{SR}} = \rho/\gamma^3 = m c/B e \gamma^2$, where $\rho$ is the bending radius of the magnetic field. Conversely, the radiation will be completely incoherent if the~bunch length is longer than $\sigma_{\textmd{SR}} N^{3/4}$. In the intermediate, partially-coherent regime, the radiated power depends critically on the current profile of the bunch, such that simulations are normally required for an accurate prediction of the effect. However, for a longitudinally Gaussian bunch we can estimate the emitted power to be
\begin{equation}
    P_{\textmd{CSR}} = \frac{\kappa e^2 c}{\epsilon_0} \frac{N^2}{\rho^{2/3} \sigma_z^{4/3}},
\end{equation}
where $\kappa \approx 0.0279$ is a numerical factor. Consider the example of a 10~GeV bunch of length 10~$\mu$m rms and charge 1~nC being bent by a magnetic field of 1~T---the radiation would be partially coherent and the bunch would radiate about 0.3\% of its energy per meter of dipole. While this is not dramatic, energy spread and chirp may be induced in a multi-dipole chicane.

Equally important to the energy loss are the \textit{transverse} CSR effects---strong intra-bunch transverse wakefields that can lead to significant emittance growth. Highly sensitive to the full 6D phase space of the bunch, these effects normally need to be studied using numerical simulations. Fortunately, with careful chicane design, transverse CSR effects can be partially mitigated \cite{JingIPAC2019}.

\subsection{Effective gradient}
With so many constraints to satisfy, how much space will be needed between stages? What is the \textit{effective gradient} of the accelerator? If, say, 10~m of space is required between 10~cm long stages operating at 5~GV/m, the effective gradient would be only 50~MV/m---not much better than a conventional machine. Although the ultimate metric will probably be the \textit{cost per collision} (linear collider) or the \textit{cost per photon} (free-electron laser), the effective gradient is a good approximate metric---it estimates the overall footprint of a machine. Staging will therefore be a crucial part of the optimization process. The current goal of the advanced accelerator community is to demonstrate an effective gradient of at least 1~GV/m \cite{CrosMuggliANAR2017}.

An interesting, and potentially worrying aspect of staged high-gradient wakefield accelerators is how the staging length scales with energy \cite{LindstromNIMA2016}. To capture and refocus the highly divergent beams, increased focusing strength will be required as the beam energy increases stage-by-stage. If we already use the strongest focusing optics available, the only solution is to make the optics \textit{longer}. How much longer? The staging length turns out to scale with $\sqrt{\gamma}$---the square root of the energy---i.e., slowly increasing with energy. This means that the effective gradient will go down as the energy goes up!

\begin{figure}[t]
	\centering\includegraphics[width=0.9\linewidth]{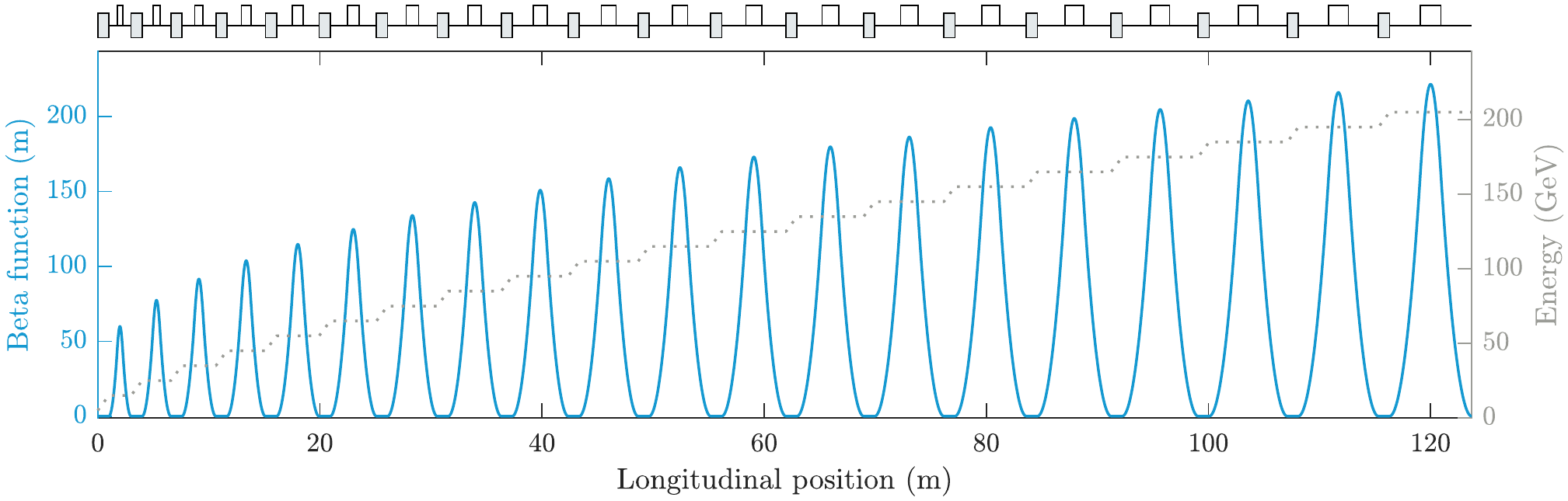}
	\caption{Illustration of how the staging length increases with energy: the accelerator stages (gray boxes) remain the~same length, whereas the length of the staging optic (white boxes) increases as $\sqrt{\gamma}$. As a consequence, the~effective gradient decreases with energy. In this example, the initial beam energy is 5~GeV, and each stage adds 10~GeV over 1~m. The beam is matched to a plasma density 10$^{16}$~cm$^{-3}$ by an optic with a magnetic field gradient 240~T/m, which takes up 25\% of the space between stages.}
    \label{fig:EnergyScaling}
\end{figure}

To understand this $\sqrt{\gamma}$-scaling, let us consider a simple beam optic between two stages (see~Fig.~\ref{fig:EnergyScaling}). Its focusing strength reduces with energy as $K \sim 1/\gamma$. Using a constant-length optic between stages would result in a focal length (and therefore capture and refocusing length) scaling of $f \approx 1/Kl \sim \gamma$. However, if the length of the optic instead scales as $l \sim \sqrt{\gamma}$, so does the focal length: $f \sim \sqrt{\gamma}$. Since both the focal length and the optic length scale similarly, the optic will always take up a constant fraction of the staging length. Furthermore, the matched beta function also scales as $\beta_m \sim \sqrt{\gamma}$, and therefore the entire optics solution (the evolution of the beta function) will scale as $\sqrt{\gamma}$---i.e., if you have found an~optics solution for one energy, you have found it for them all.

%--- Figure: Energy scaling ---%

\section{Proposed Methods}
\label{sec:ProposedMethods}
Faced with all the above challenges and requirements, it is clear that we need to innovate in order to succeed. Several new ideas are attempting to either tackle the problems head on, or to circumvent them by avoiding staging altogether. This section will go through a few of the most notable proposals so far.

\subsection{Plasma density ramps}
\begin{figure}[ht]
	\centering\includegraphics[width=0.75\linewidth]{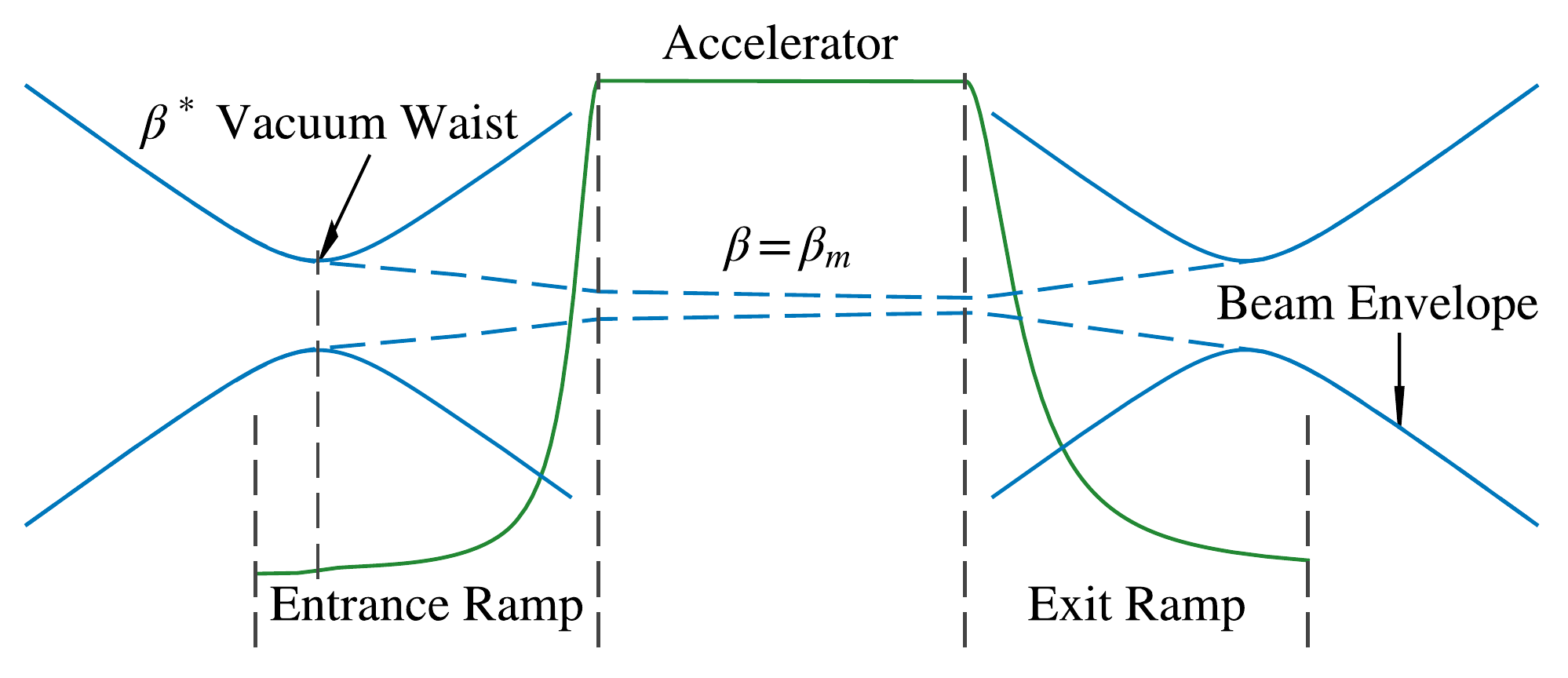}
	\caption{Schematic of a plasma accelerator with density ramps. The beam is focused by external focusing to the~vacuum waist ($\beta^*$) close to the start of the entrance ramp. If perfectly matched to the ramp, the beam stays matched throughout the accelerator ($\beta = \beta_m$ in the flat-top). Finally, the beam is transported through the exit ramp, which reduces the divergence before exiting into the outside vacuum. Source: R.~Ariniello \textit{et al.}, Phys.~Rev.~Accel.~Beams \textbf{22}, 041304 (2019) \cite{ArinielloPRAB2019} (CC BY 4.0).}
    \label{fig:PlasmaDensityRamp}
\end{figure}

An obvious solution to the strong-focusing conundrum is to simply reduce the strength of the focusing and thereby increase the matched beta function---in a plasma accelerator, this corresponds to reducing the plasma density (see Eq.~(\ref{eq:MatchedBeta})). This mitigates the divergence/chromaticity problem, but comes at the~cost of reducing the accelerating gradient. However, the plasma density does not need to be the same everywhere---we can use a higher density throughout most of the stage for high gradient acceleration, and a lower density at the entrance and exit for reduced beam divergence. This longitudinal density tailoring is often called a \textit{plasma density ramp} \cite{MarshPAC2005,DornmairPRAB2015,ArinielloPRAB2019}. Figure~\ref{fig:PlasmaDensityRamp} illustrates this concept.

%--- Figure 7: Plasma density ramp ---%
\newpage
Calculating the evolution of the beta function through a tailored plasma density profile is relatively straightforward via the \textit{betatron equation} \cite{MartiniCERNPS1996}
\begin{equation}
    \label{eq:Betatron}
    \frac{1}{2} \beta''(s) \beta(s)  - \frac{1}{4} \beta'(s)^2 + K(s) \beta(s)^2 = 1,
\end{equation}
where the focusing force is given by
\begin{equation}
    K(s) = \frac{e^2}{2 \epsilon_0 } \frac{n(s) }{E(s)}.
\end{equation}
Note that both the exposed charge density $n(s)$ and the particle energy $E(s)$ are changing with the~longitudinal position $s$. While it is possible to solve Eq.~(\ref{eq:Betatron}) analytically in certain cases \cite{XuPRL2016}, in general it needs to be integrated numerically.

It is important to note that although plasma density ramps reduce the divergence, they do not necessarily solve the chromaticity problem---while one energy slice might be matched and emittance-preserved, this does not mean that every energy slice will. To ensure that all energies remain matched throughout the accelerator, the ramps must be \textit{adiabatic} \cite{HelmSLAC1962,FloettmannPRAB2014}. This effectively means that the plasma density is changing sufficiently slowly,
\begin{equation}
    \left|\frac{n'(s)}{n(s)}\right| \ll \frac{1}{\beta_m(s)},
\end{equation}
such that $\alpha \approx 0$ throughout the entire ramp. An example of such a ramp would be $n(s) = n_0 (1+s/l_r)^{-2}$, where $n_0$ is the flat-top density and $l_r \gg 2\beta_{m0}$ is the characteristic ramp decay length ($\beta_{m0}$ is the~matched beta function in the flat-top). To reach a beta function $\beta^*$ at the entrance/exit of such an~adiabatic ramp, the overall ramp length must be $L_r \gg 2 \beta^*$.

Adiabatic ramps are desirable due to their insensitivity to energy spread or slight mismatching, but this comes at the price of significantly longer ramp sections. Long ramps will both reduce the energy efficiency and the effective gradient of the accelerating structure, and can introduce potentially non-negligible and non-uniform decelerating fields, which must be compensated for in the main accelerating section. In short, plasma density ramps constitute a crucial tool for reducing high divergence and chromatic effects, but will likely not be able to entirely solve the problem.

\subsection{Plasma lenses}
Another useful tool for capturing and refocusing beams between stages is the \textit{plasma lens}---a charged particle optics device that provides strong focusing in both planes simultaneously (as opposed to the~quad-rupole). Plasma lenses fall into two categories: \textit{passive} and \textit{active} plasma lenses, referring to whether the focusing force is externally (actively) driven or not.

\subsubsection{Passive plasma lenses}
\begin{figure}[t]
	\centering\includegraphics[width=0.6\linewidth]{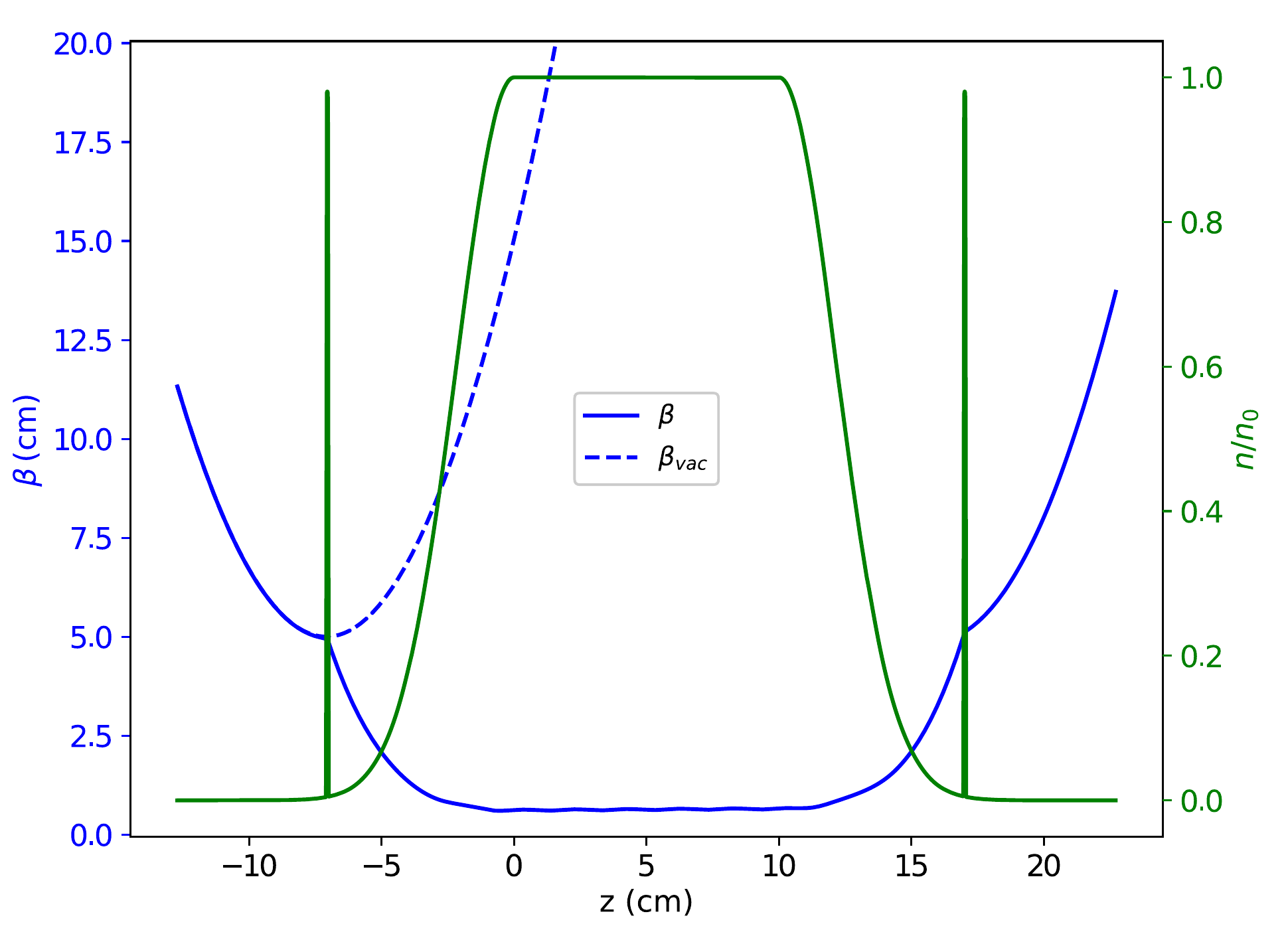}
	\caption{Passive plasma lensing scheme proposed for FACET-II \cite{JoshiPPCF2018}, where the lenses are integrated into the plasma density ramps. A gas jet produces a locally higher gas density that is ionized by a laser. The lens is driven by the~same beam driver as is used in the main accelerating section. Source: C.~E.~Doss \textit{et al.}, Phys.~Rev.~Accel.~Beams \textbf{22}, 111001 (2019) \cite{DossPRAB2019} (CC BY 4.0).}
    \label{fig:PassivePlasmaLens}
\end{figure}

Passive plasma lenses utilize the same mechanism as plasma density ramps---plasma wakefield focusing. The wakefield can either be driven by the beam itself or by a separate driver (laser or particle beam). Typically, such lenses can provide very strong focusing fields---in the MT/m-range---and so can be made very compact. While the concept dates all the way back to 1922 \cite{JohnsonJOSA1922,BorriesZP1932,BennettPR1934,GaborNature1947}, passive plasma lenses in their modern form were proposed in 1989 \cite{ChenPRD1989} and have been successfully demonstrated for both beam drivers \cite{RosenzweigPFB1990,NgPRL2001} and laser drivers \cite{Thaury2015}. Figure~\ref{fig:PassivePlasmaLens} illustrates how a passive plasma lens might be applied in practice.

%--- Figure: Passive plasma lens ---%

Two regimes are often identified: the \textit{underdense} and the \textit{overdense} regime, referring to whether the plasma density in the lens is lower or higher than the beam density, respectively. If the lens is underdense, a nonlinear plasma wakefield (a blowout) forms with a fully exposed ion column---this provides a linear focusing force and is therefore in principle emittance preserving. Although the focusing force for electrons is exerted by an electric field, we can calculate the equivalent magnetic field gradient (as $E_r \equiv c B_{\phi}$ for ultra-relativistic particles)
\begin{equation}
    \label{eq:PassivePlasmaLensing}
    g_{\textmd{PPL}} = \frac{e n}{2 c \epsilon_0},
\end{equation}
where $n$ is the plasma density of the lens. This focusing is uniform for all particles inside the wake.

Compare that to a lens in the overdense regime, where a linear wakefield forms---in this case the~\textit{local} focusing force will also be given by Eq.~(\ref{eq:PassivePlasmaLensing}), but the exposed charge density can vary throughout the beam, both transversely and longitudinally. This nonuniform focusing force can result in emittance growth. The same effect can also occur in underdense plasma lenses for self-focused beams, where the~wake builds up longitudinally along the bunch, resulting in a projected emittance growth. 

Note that since the lenses are usually very short, energy changes from the longitudinal wakefield are typically ignored.

\subsubsection{Active plasma lenses}
An alternative way to focus beams is to use the plasma as a conductor, and use large currents to produce strong magnetic fields. This actively-driven plasma lens can provide a uniform focusing field for the~entire bunch without the need for a driver, which make them compact and simple to operate. On the~other hand, the focusing strength is typically limited to the kT/m-range---orders of magnitude weaker than passive plasma lenses, but still very strong compared to conventional quadrupoles. This is the type of plasma lens that was used for the BELLA staging experiment (see Fig.~\ref{fig:BELLAstaging}) \cite{SteinkeNature2016}.

\begin{figure}[t]
	\centering\includegraphics[width=0.5\linewidth]{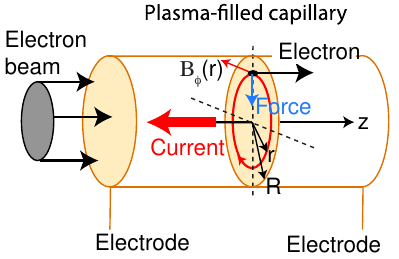}
	\caption{Schematic of an active plasma lens: an electron beam enters a gas-filled capillary, which is discharged via two high-voltage electrodes. The current flowing between the electrodes induces an azimuthal magnetic field $B_{\phi}$ that increases with radius $r$---a radial focusing force that focuses the electron beam in both planes. Source: J.~van Tilborg \textit{et al.}, Phys.~Rev.~Lett.~\textbf{115}, 184802 (2015) \cite{vanTilborgPRL2015} (reproduced with permission).}
    \label{fig:ActivePlasmaLens}
\end{figure}

The history of active plasma lenses started in 1950 \cite{PanofskyRSI1950,ForsythIEEE1965}, during which they were studied for various purposes such as ion focusing \cite{BoggaschPRL1991} and antimatter capture \cite{KowalewiczPAC1991} before their recent revival for use in plasma accelerators \cite{vanTilborgPRL2015}. Modern active plasma lenses consist of a thin (mm-scale) gas-filled capillary with electrodes on either side \cite{LindstromNIMA2018} (see Fig.~\ref{fig:ActivePlasmaLens} for a schematic overview). A high-voltage discharge ionizes the gas before a large current passes through the plasma. By Ampere's law, the azimuthal magnetic field at each radius inside the lens is proportional to the total current enclosed at that radius. If, ideally, the current density inside is uniform, the resulting magnetic field is linear and has a field gradient
\begin{equation}
    \label{eq:ActivePlasmaLensing}
    g_{\textmd{APL}} = \frac{\mu_0 I}{2\pi R^2},
\end{equation}
where $I$ is the overall current, $R$ is the capillary radius and $\mu_0$ is the permittivity of free space. As an example, plugging in numbers for a typical lens of radius 500~$\mu$m and current 500~A, we obtain a~magnetic field gradient of 400~T/m---a relatively strong lens. Equation~(\ref{eq:ActivePlasmaLensing}) also indicates an advantage of active plasma lenses: by inverting the direction of the current, the lens can also focus positively charged particles---this is not easily done with a passive plasma lens.

%--- Figure: Active plasma lens ---%

Unfortunately, active plasma lenses are not always emittance preserving. There are three principal ways in which these lenses can degrade the beam quality:
\begin{enumerate}

    \item \textit{Nonuniform current density} \cite{vanTilborgPRAB2017} leads to nonlinear focusing fields, which causes emittance growth. This can be caused by a temperature gradient between the core and the wall (where the heat escapes) \cite{BobrovaPRE2001,BroksPRE2005}. Interestingly, while this aberration is present in light gases like hydrogen and helium \cite{RockemannPRAB2018}, it is possible to fully suppress it in a heavier gas like argon where the heat transfer to the wall is significantly slower \cite{LindstromPRL2018}. Another effect which can cause nonuniform focusing fields is the \textit{z-pinch} effect \cite{ChristiansenCERNPS1984}, where the magnetic field of the lens is strong enough to self-focus its own current.
    
    \item \textit{Coulomb scattering} \cite{WiedemannSpringer2007,KirbyPAC2007} can cause emittance growth due to the atomic/ionic density on axis. Since the emittance growth rate scales as
    \begin{equation}
        \frac{d \epsilon_n}{ds} \sim \frac{n \beta Z^2}{\gamma},
    \end{equation}
    where $Z$ is the atomic number and $\beta$ is the beta function in the lens, scattering is much more severe for heavier gases---the effect is usually negligible in hydrogen and helium, but can be a problem for an argon-based active plasma lens.
    
    \item \textit{Passive plasma lensing} will also occur in an active plasma lens if the beam density is sufficiently high---which it often is if the lens is placed close to the exit of a wakefield accelerator. Currently, this appears to be the main limiting factor for the application of active plasma lenses to beams relevant to FELs or linear colliders \cite{LindstromArXiv2018}.
\end{enumerate}
Clearly, any application of active plasma lensing needs to take all the above effects into account.

\subsection{Achromatic beam transport}
Chromaticity in accelerators is nothing new---several strategies for mitigating it exist. In particular, collider final focusing shares many of the same challenges---how to deal with highly diverging/converging beams with non-negligible energy spread---and therefore much thought has already gone into solving these issues. The bottom line is that even though beam transport with a single focusing optic is chromatic, we can often construct lattices of multiple elements that are effectively \textit{achromatic}.

\subsubsection{Apochromatic correction}
\begin{figure}[ht]
	\centering\includegraphics[width=0.7\linewidth]{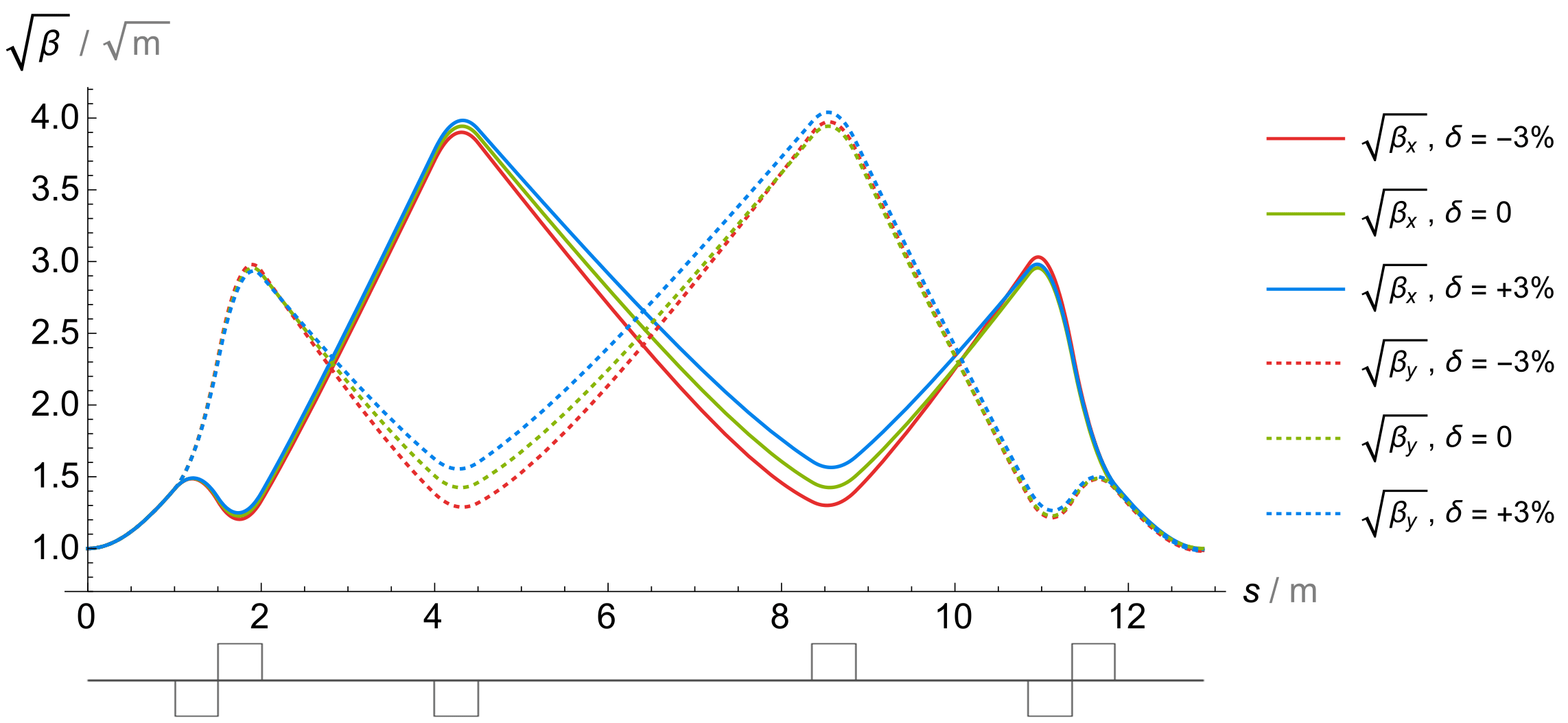}
	\caption{Illustration of apochromatic correction in a lattice of only linear focusing optics (six quadrupole magnets). The lattice is constructed such that beta functions evolve differently for the various energy slices, but eventually converge before entering the next stage. This method also works for plasma lenses and plasma density ramps. Source: C.~A.~Lindstr{\o}m and E. Adli, Phys.~Rev.~Accel.~Beams \textbf{19}, 071002 (2016) \cite{LindstromPRAB2016} (CC BY 3.0).}
    \label{fig:Apochromat}
\end{figure}

Using linear optics elements, it is fundamentally impossible to provide fully achromatic beam transport for all energies \cite{WolskiICP2014}. However, it \textit{is} possible to cancel chromaticity for a limited energy spread at certain locations in the lattice. This is inspired by how focusing of multiple colors is done in camera lenses: introduce more degrees of freedom (i.e., more lenses) and tune the system such that several colors are all in focus. In beam optics, this corresponds to different energies traversing the lattice with different beta function evolution, but eventually converging to the same focus (see Fig.~\ref{fig:Apochromat}). This idea was first introduced for the final focusing of CLIC in 1987, and is known as \textit{aprochromatic correction} \cite{MontagueCLIC1987}.

%--- Figure: Apochromatic focusing ---%

A beam transport lattice can be apochromatically corrected to arbitrary order \cite{LindstromPRAB2016}, such that not only the first order chromatic amplitude is zero, but also higher orders---at the cost of introducing more degrees of freedom (optics elements). However, this achromatic behavior applies only to a limited range of energies around the nominal energy. This transportable energy range is roughly $\sigma_{\delta} \approx 1/W$, where $W$ is the chromatic amplitude of a more basic lattice where no apochromatic correction is applied. Beyond this characteristic energy spread, full emittance preservation is not possible. Therefore, apochromatic correction should be used in combination with other methods that reduce the intrinsic chromaticity---such as plasma density ramps and plasma lenses.

\subsubsection{Sextupoles in dispersive sections}
In order to truly increase the energy acceptance of a beamline, nonlinear optics must be introduced. The~conventional solution is to use sextupole magnets in regions of large dispersion. This is the solution employed in collider final focusing, where the chromaticity is so large ($W \gtrsim 10^4$) that apochromatic focusing is incapable of correcting for the required energy spread ($\sigma_{\delta} \approx 1\%$). Many high-gradient wakefield accelerators have similar parameters, and may therefore require nonlinear optics lattices.

\begin{figure}[ht]
	\centering\includegraphics[width=0.7\linewidth]{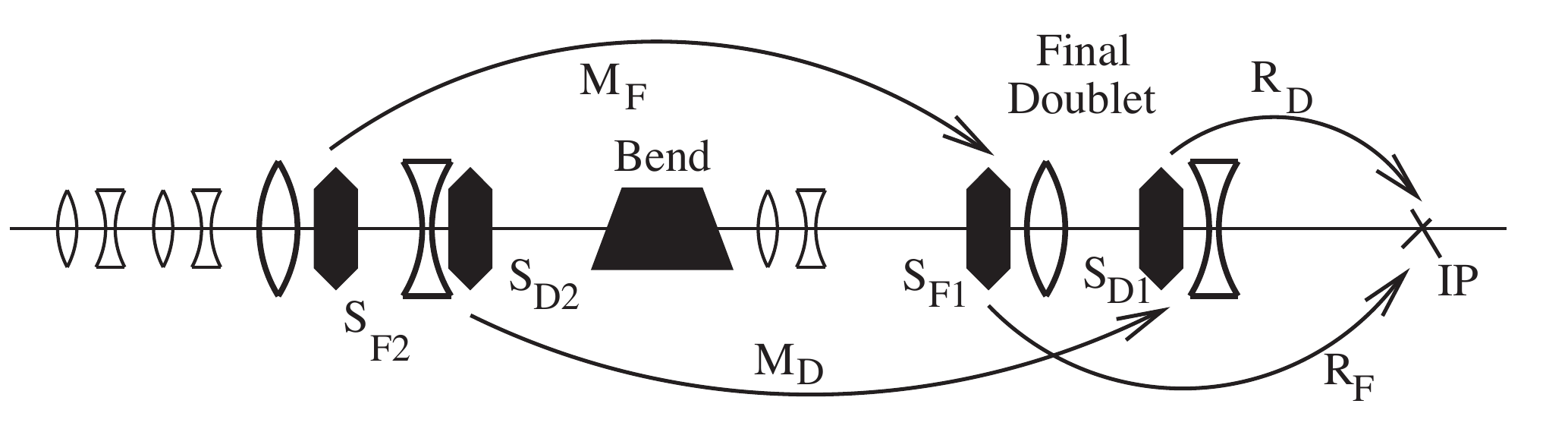}
	\caption{Local chromaticity correction for a linear collider using nonlinear optics and dispersive elements. The~beam is dispersed with a dipole (black trapezoid) onto the final doublet of quadrupoles (white). Each quadrupole has a corresponding sextupole (black hexagons) to locally correct the chromaticity. Prior to the dipole is a similar doublet that compensates the geometric terms introduced by each of the two final sextupoles. Source: P.~Raimondi and A.~Seryi, Phys.~Rev.~Lett.~\textbf{86}, 3779 (2001) \cite{RaimondiPRL2001} (reproduced with permission).}
    \label{fig:LocalChromCorr}
\end{figure}

The underlying concept of chromaticity correction with sextupoles is relatively straightforward. The local focusing fields in a sextupole are proportional to the transverse offset, so if the beam is dispersed such that different energies enter the sextupole at different offsets, the chromaticity can be cancelled. If the beam is horizontally dispersed ($x \to x + D_x \delta$), the focusing forces are given by
\begin{eqnarray}
    F_x &\sim& x D_x \delta + \frac{1}{2}\left(x^2-y^2\right) + \frac{1}{2}D_x^2 \delta^2 \\
    F_y &\sim& y D_x \delta + yx
\end{eqnarray}
for the energy slice $\delta$, where the first terms ($x\delta$ and $y\delta$) can be used for chromaticity-correction, and the rest are nonlinear geometric ($x^2$, $y^2$ and $yx$) and chromatic terms ($\delta^2$). These nonlinear forces cause emittance growth---an effect that must be mitigated by introducing another sextupole elsewhere in the~lattice to exert exactly the opposite nonlinear forces. See Fig.~\ref{fig:LocalChromCorr} for an illustration of this concept.

A particularly important concept for mitigating chromaticity is that of \textit{local chromaticity correction} \cite{RaimondiPRL2001,WhitePRL2014}. While it is possible to correct chromaticity \textit{globally} by distributing sufficiently many sextupoles across the lattice to be able to cancel chromaticity at the end \cite{IrwinIEEE1991}, it will always be better to correct the chromaticity \textit{locally} by placing a dedicated sextupole next to each chromaticity-inducing focusing optic. This greatly increases the energy acceptance of the lattice, because chromaticity never really develops in the first place. All modern collider designs \cite{ILCTDR2013,CLICCDR2013} employ local chromaticity correction in their final-focusing systems.

%--- Figure: Local chromaticity correction ---%

The drawbacks of chromaticity correction using nonlinear optics are: (1) the introduction of large dispersion, and (2) long and complex lattices. Installing two collider-style final-focusing systems back-to-back between the stages of a wakefield accelerator would take up a large amount of space---defeating the purpose of the high-gradient acceleration. However, if local chromaticity correction can be applied in a simpler and more optimized scheme (e.g., utilizing plasma lenses and mirror symmetry), this might hold the key to compact and chromaticity-free staging.

\subsection{Single-stage acceleration}
Sometimes, the only way to win is not to play. We should therefore briefly consider some alternative methods for accelerating to high energies without using multiple stages---i.e., \textit{single-stage acceleration}. Before delving in, it is important to note that all these techniques will also require some form of emittance-preserving out-coupling---which means that many of the above considerations are still relevant, but the requirements for compactness can be relaxed.
\newpage
\subsubsection{Proton-driven plasma accelerators}
The main hurdle to achieving single-stage wakefield acceleration to high energy is the overall energy content of the driver---it needs to be very high. While lasers and electron beams rarely go beyond 100~Joules, large synchrotrons can provide proton beams with 1--100~kJ of energy per bunch---more than enough to accelerate 1~nC to 1~TeV. Unfortunately, these proton bunches are not short enough to drive high-frequency, high-gradient wakefield accelerators. The solution is to transform the long proton beam into a train of many short bunches, and let the wakefield build up resonantly along the train. Interestingly, this can be done in a plasma using a process known as \textit{self-modulation} \cite{KumarPRL2010,SchroederPRL2011}.

\begin{figure}[ht]
	\centering\includegraphics[width=0.85\linewidth]{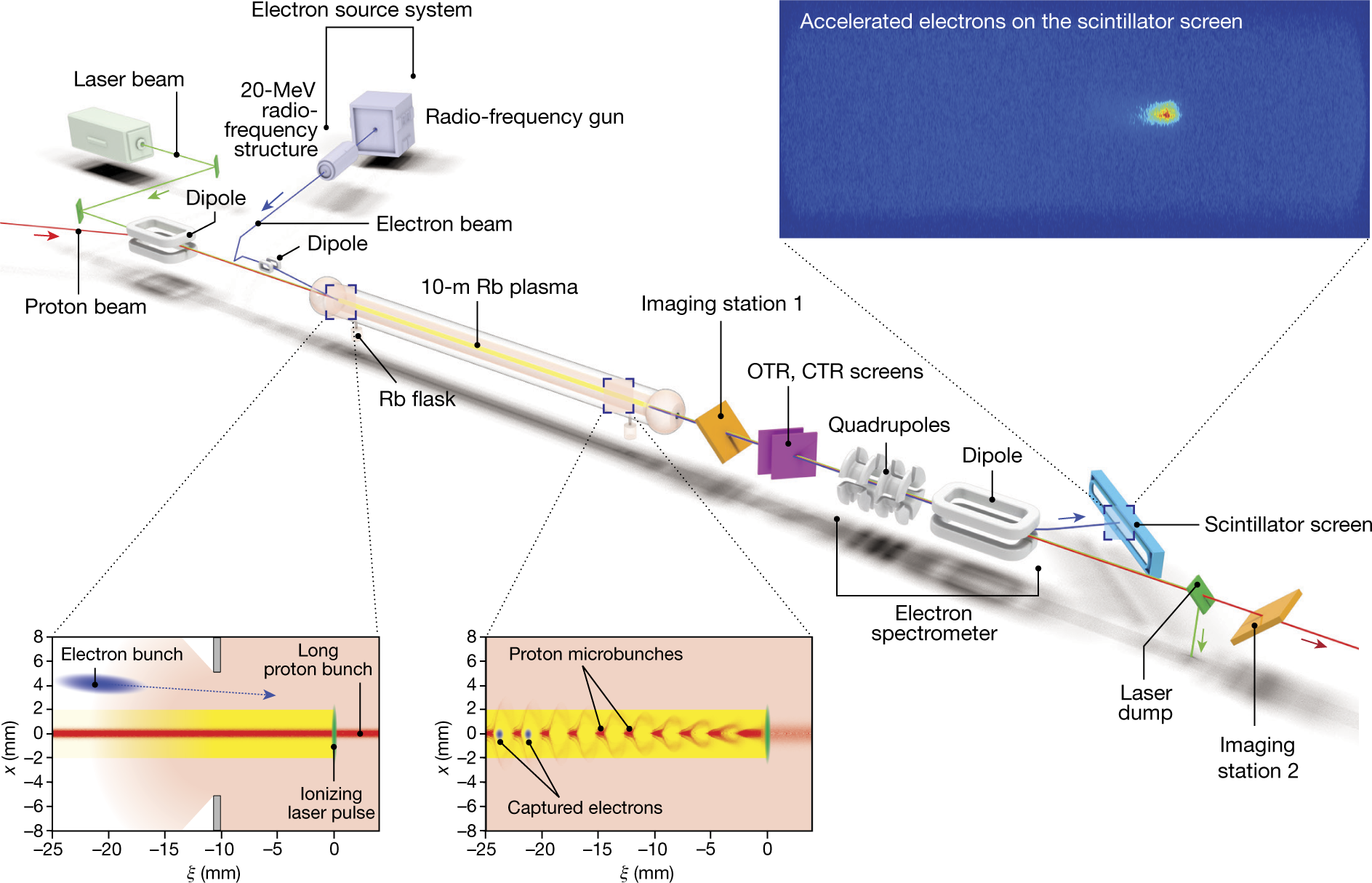}
	\caption{The AWAKE experimental setup, using a proton bunch in a laser-ionized plasma to accelerate an electron beam to 2~GeV (upper right). The long proton bunch self-modulated (lower left) into a train of short bunches, resonantly driving a plasma wakefield, into which electrons were externally injected and then accelerated. Source: E.~Adli \textit{et al.}~(AWAKE Collaboration), Nature \textbf{561}, 363 (2018) \cite{AWAKENature2018} (CC BY 4.0).}
    \label{fig:AWAKE}
\end{figure}

Self-modulation is a process whereby a beam in a plasma self-interacts with its own focusing and defocusing wakefields. Where the beam is focused it gets denser; where the beam is defocused it gets ejected---amplifying the on-axis charge modulation, and increasing the amplitude of the wakefield. This instability gradually builds up until the proton beam is fully self-modulated. The AWAKE experiment \cite{GschwendtnerNIMA2016} at CERN, utilizing 400~GeV proton bunches from the Super Proton Synchrotron has successfully demonstrated this self-modulation \cite{AWAKEPRL2019,TurnerPRL2019}, and used it to accelerate electron bunches up to 2~GeV \cite{AWAKENature2018} in a 10~m long plasma stage (see Fig.~\ref{fig:AWAKE}).

%--- Figure: AWAKE experiment ---%

While self-modulated proton-driven wakefield acceleration constitutes one of the most promising ways of realizing high-gradient, high-energy acceleration, there are also a few limitations. Firstly, a very large proton synchrotron is required---it may only make sense to build such accelerators in the vicinity of existing infrastructure. The limited repetition rate of such machines is also of concern. Secondly, the external injection into the wakefield is non-trivial---it will be particularly challenging to preserve the~emittance of the electron bunch. Moreover, if a separate proton self-modulator stage is required before the injection and acceleration of electrons, many of the above-mentioned staging challenges will apply.

\subsubsection{Travelling-wave electron acceleration}
\begin{figure}[t]
	\centering\includegraphics[width=0.7\linewidth]{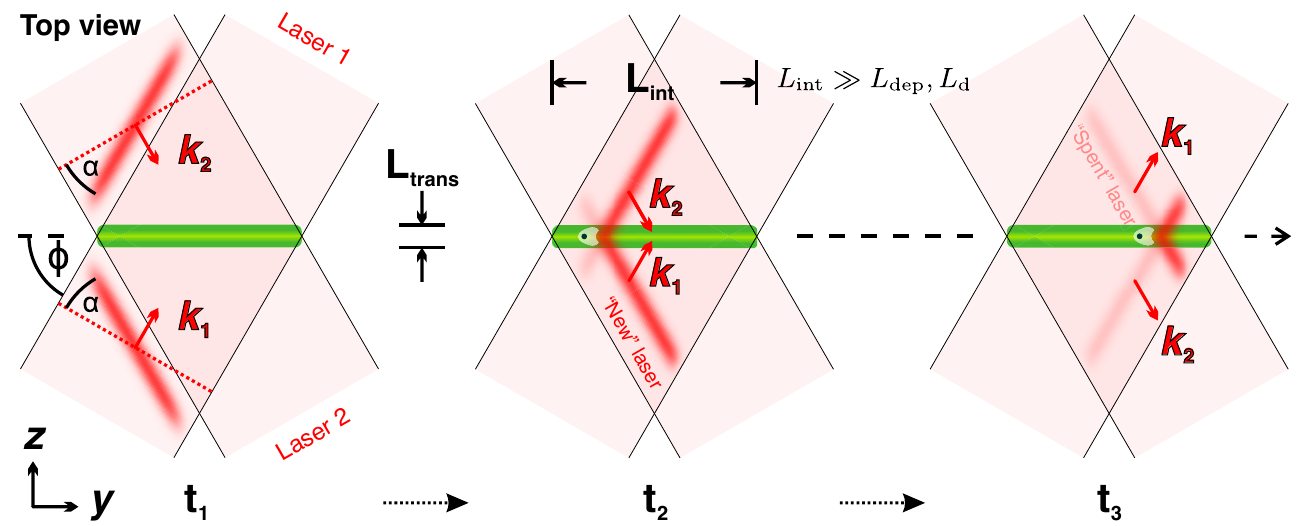}
	\caption{Illustration of the travelling-wave electron acceleration scheme: two lasers are coupled in transversely into an on-axis plasma channel. The superposition of the two lasers drives a wakefield, and since the laser pulse fronts are tilted, the travelling wave can be made to move at speed $c$. Source: A.~Debus \textit{et al.}, Phys.~Rev.~X \textbf{9}, 031044 (2019) \cite{DebusPRX2019} (CC BY 4.0).}
    \label{fig:TWEAC}
\end{figure}

A proposal for reaching high energies in a single-stage laser-driven wakefield accelerator is using a~\textit{travelling} laser focus. The idea is to couple in two laser drivers transversely---one from each side---and use their superposition to drive the wakefield. Further, if their pulse fronts are tilted just right, the laser focus can be made to travel at the speed of light $c$---to avoid dephasing between the laser and the electron beam. Figure~\ref{fig:TWEAC} illustrates this so-called travelling-wave electron acceleration scheme \cite{DebusPRX2019}.

%--- Figure: Travelling wave electron acceleration ---%

The scheme is interesting because it exploits the ability of lasers to be transversely in-coupled, and may allow seamless blending of multiple lasers with no need for staging. On the other hand, it will require an extreme level of control over the laser parameters that may be difficult to achieve in practice.

\subsubsection{Curved plasma channels}
\begin{figure}[h]
	\centering\includegraphics[width=0.75\linewidth]{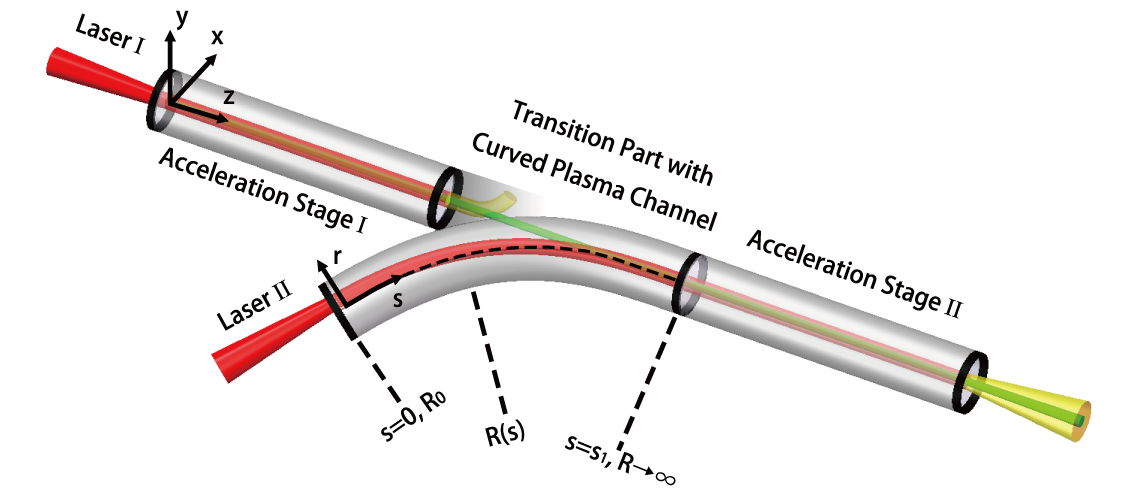}
	\caption{Illustration of the curved plasma channel scheme: after an initial laser plasma accelerator stage, a fresh laser pulse (red) is in-coupled using a curved plasma channel. This channel guides the laser around the curve, but lets the electron beam (green) pass straight through until it is trapped by the new wakefield. The depleted laser (yellow) from the first stage is also out-coupled in this transition region. Source: J.~Luo \textit{et al.}, Phys.~Rev.~Lett.~\textbf{120}, 154801 (2018) \cite{LuoPRL2018} (reproduced with permission).}
    \label{fig:CurvedChannels}
\end{figure}

Finally, another laser-based method has been proposed for swapping out the driver without staging: coupling in fresh laser drivers directly using \textit{curved plasma channels} \cite{LuoPRL2018}. Similar to how laser pulses are guided in straight plasma channels with a transverse (parabolic) density profile \cite{DurfeePRL1993,SpencePRE2000}, a laser pulse can also follow a curved channel. In contrast, a high-energy electron bunch will pass straight through such a plasma profile, and can therefore be handed off from one driver to the next with minimal disruption. Figure~\ref{fig:CurvedChannels} illustrates this scheme.

%--- Figure: Curved plasma channels ---%

While the beam never really exits the plasma during acceleration, transitioning from one stage to the next will introduce transverse oscillations to the beam centroid. This causes emittance growth and may lead to a beam-breakup instability. Nevertheless, mitigation strategies may be found---in which case this scheme promises an achievable path towards high energy acceleration with high-gradient wakefields.
\pagebreak
\section{Conclusions}
The application of high-gradient wakefields to truly high-energy accelerators is currently held back by the practical problems of staging. In the conventional view, the driver must be in- and out-coupled in the~space between stages. Combined with the strong focusing and finite energy spread of such accelerators, the capture and refocusing of beams become highly chromatic---resulting in large emittance growth and potentially beam loss. For the beam quality to remain high throughout a multi-stage accelerator, many requirements must be met: matching of beta functions, dispersion cancellation, isochronicity, avoiding too much synchrotron radiation, as well as tight synchronization and misalignment tolerances. At the~same time, the whole setup needs to be compact to ensure a high effective acceleration gradient. 

Many good ideas have been proposed for how to mitigate or avoid the challenges of staging. The~concept of plasma density ramps is key to reducing the high divergence---and if made adiabatic, they can help reduce the chromaticity. Passive and active plasma lenses can take the next step: capturing and refocusing the beams more compactly than is possible with conventional quadrupoles. Nevertheless, specialized beam transport optics will likely still be required to handle the residual chromaticity---this can either be done with the use of apochromatic staging, or better yet: nonlinear optics with local chromaticity correction. Finally, alternative methods have been proposed that sidestep staging altogether by employing single-stage acceleration. Examples include self-modulated proton-driven plasma wakefields, laser-based travelling-wave electron acceleration, and laser-driven acceleration in curved plasma channels. 

In summary, staging of high-gradient accelerators is still a young field, and the best solutions are surely still to come!

\section*{Acknowledgements}
The author wishes to thank E.~Adli for input to the CERN Accelerator School lecture, on which this publication is based.

\end{document}